# Controlling in-gap end states by linking nonmagnetic atoms and artificially-constructed spin chains on superconductors

Lucas Schneider[1,†], Sascha Brinker[2,3,†], Manuel Steinbrecher[1,‡], Jan Hermenau[1], Thore Posske[4], Manuel dos Santos Dias[2], Samir Lounis[2], Roland Wiesendanger[1], and Jens Wiebe[1,*]

[1]Department of Physics, Universität Hamburg, D-20355 Hamburg, Germany.

[2]Peter Grünberg Institute and Institute for Advanced Simulation, Forschungszentrum Jülich & JARA, D-52425 Jülich, Germany.

[3]Department of Physics, RWTH Aachen University, 52056 Aachen, Germany

[4]I. Institute for Theoretical Physics, Universität Hamburg, D-20355 Hamburg, Germany.

[‡]Present address: Institute for Molecules and Materials (IMM), Radboud University, Nijmegen, The Netherlands.

[†]These authors contributed equally to this work.

[*]E-mail: jwiebe@physnet.uni-hamburg.de

**Chains of magnetic atoms with either strong spin-orbit coupling or spiral magnetic order which are proximity-coupled to superconducting substrates can host topologically non-trivial Majorana bound states[1–7]. The experimental signature of these states consists of spectral weight at the Fermi energy and spatially localized near the ends of the chain[8–13]. However, topologically trivial Yu-Shiba-Rusinov in-gap states localized near the ends of the chain can lead to similar spectra[9,14–17]. Here, we explore a protocol to disentangle these contributions by artificially augmenting a candidate Majorana spin chain[11] with orbitally-compatible nonmagnetic atoms. Combining scanning tunneling spectroscopy with ab-initio and tight-binding calculations, we realize a sharp spatial transition between the proximity-coupled spiral magnetic order and the non-magnetic superconducting wire termination, with persistent zero-energy spectral weight localized at either end of the magnetic spiral. Our findings open a new path towards the control of the spatial position of in-gap end states, trivial or Majorana, via different chain terminations, and the realization of designer Majorana chain networks for demonstrating topological quantum computation.**

**Main**

Candidate Majorana platforms based on magnetic chains have been experimentally realized by self-assembled growth of chains of Fe[8–10,12,13] or Co atoms[14] on superconducting Pb(110), and by the controlled atom-by-atom assembly of Fe chains on superconducting Re(0001) using the tip of a scanning tunneling microscope (STM)[11]. For both substrates, scanning tunneling spectroscopy (STS) supplied experimental evidence for topological superconductivity in Fe chains via the observation of zero-energy spectral weight localized at the ends of such chains, which was interpreted as the signature of a Majorana bound state[18,19]. For the Pb system[12], it was shown theoretically that the Majorana bound state can have a strong spectral weight in the superconducting substrate close to the chain's end, which was experimentally supported by the detection of zero-energy spectral weight localized in Pb overlayers covering the Fe chains close to their ends. However, for all investigated systems so far, the experiments suffer from a concurrence of the expected location of the Majorana bound state with the location of the termination of the chain. This termination is intrinsically different from the inner part of the chain, as the end atoms have a different coordination[17]. Thereby, these atoms tend to have a different bonding length and, thus, a different hybridization with the substrate as compared to those in the interior of the chain. Moreover, the spin-related properties at the ends of magnetic chains can differ drastically from their interior[20–22]. All this could lead to a localization of topologically trivial Yu-Shiba-Rusinov (YSR) states at the chain's ends and also accidentally close to the Fermi energy, and, therefore, hinder the identification of the spectral weight stemming from the sought-after Majorana bound states. A solution to these issues would be the termination of the magnetic chain by a nonmagnetic superconducting wire made from a material with a similar orbital structure: such a termination ensures a smooth electronic continuation of the magnetic chain, which we expect to have a stronger impact on the properties of the YSR states than on those of the Majorana bound state, due to the robustness inherent to the topological character of the latter. In order to investigate the possibilities to realize such terminations, we study artificial chains of three different species from the *3d* transition metals series (Mn, Fe, Co), as well as hybrid chains of Co and Fe atoms, assembled by STM-tip induced manipulation on superconducting Re(0001).

By depositing the three transition metal elements onto the cold Re(0001) substrate and subsequent STM-tip induced manipulation (see Methods), we place single Fe and Co atoms on two different adsorption sites: the hollow site which continues the hexagonal close packed (hcp) stacking of the substrate, and the hollow site which corresponds to the stacking of a face centered cubic (fcc) crystal. For single Mn atoms, only the fcc site is accessible[23]. Due to increasing Kondo coupling with increasing *d*-state filling and a transition from out-of-plane to easy-plane magnetic anisotropy, the energy of the YSR states which the five species induce in the energy gap of the superconducting substrate vary systematically from $Mn^{fcc}$ over $Fe^{fcc}$, $Fe^{hcp}$, $Co^{fcc}$ to $Co^{hcp}$: the YSR states of $Mn^{fcc}$ are located close to the superconducting substrate's gap edge, the ones of $Fe^{hcp}$ close to the center of the gap and the ones of $Co^{fcc}$ again at the gap edge. Interestingly, single $Co^{hcp}$ atoms are found to have fully quenched magnetic moments and thus do not induce any in-gap state[23]. Moreover, artificial chains of $Fe^{hcp}$ atoms on neighboring sites (Fig. 1c) form spin spirals and reveal strong indications for topological superconductivity by zero-energy spectral weight localized at the ends of chains which are longer than 10 atoms[11]. Motivated by these previous results, we explore the possibilities to build chains of the other two elements, Mn and Co, on neighboring sites using STM-tip induced manipulation (Fig. 1a,b).

We find that it is possible to manipulate straight chains of Co atoms on neighboring hcp sites (Fig. 1b) and zig-zag-shaped chains of Mn atoms adsorbed on neighboring sites alternating between fcc and hcp (see Fig. 1a, Supplementary Note 1, and Supplementary Fig. 1). The chains are up to more than 100 atoms long limited by the widths of the terraces of the substrate, residual substrate defects, and the number of available single atoms in the surrounding of the building area. However, it was impossible to build any other chain of closed packed atoms of one of these three elements, e.g.

straight chains of Mn atoms on neighboring fcc or hcp sites, or straight chains of Co atoms on neighboring fcc sites. This is most probably a result of an energetically unfavorable bond length in such configurations.

Next, we study the low-energy electronic properties of three manipulated chains, as shown in Fig. 1, in the energetic region of the energy gap $2\Delta = 0.51$ meV of the superconducting substrate (Fig. 2). The spectral intensity (Figs. 2b,f,j) is symmetric with respect to the center of the chain. This is in particular true for the ends of the chains (see Fig. 4b for Fe and Supplementary Note 2 and Supplementary Fig. 2 for Mn). The $Fe_{20}$ chain (Fig. 2f) reveals a zero-energy spectral weight with a maximum localized on the two atoms that terminate the chain, which decays in an oscillatory fashion in intensity towards the center of the chain. Spectra taken at the ends of the chain in comparison to spectra taken at the center of the chain (Figs. 2g,h) show, that, in addition to this zero-energy spectral weight, also the spectral intensity stemming from YSR bands at a non-zero energy of about +0.1 meV is increased towards the chain's ends (see arrows in Figs. 2f,g). This reproduces the data of our previous publication, which was taken in a different STM facility using a different STM-tip and sample[11]. With the help of ab-initio and tight-binding model calculations, the zero-energy spectral weight was interpreted as a signature for a Majorana bound state localized at each end of the $Fe_{20}$ chain.

In stark contrast, the $Mn_{101}$ chain's ends do not show any zero-energy spectral weight (Fig. 2b). In the interior of the chain, there is a YSR band whose energy is slightly smaller than $\Delta$, which is visible as a shoulder of the coherence peak on the negative bias side (Figs. 2c,d). These results let us conclude, that the relatively weak Kondo coupling[23] of Mn compared to Fe prevents the development of a topologically superconducting phase via the YSR bands[4]. Notably, the energy of the YSR band is slightly decreased at the chain's ends, and thus somewhat approaches the Fermi level (see arrows in Figs. 2b,c). The changes to the YSR band close to the ends of the Fe chain (see above) and to the YSR band energy close to the ends of the Mn chain could be explained by the reduced coordination number of the chain-terminating atoms. Such changes can, therefore, be reduced by attaching nonmagnetic atoms with a similar orbital structure as the ones in the magnetic chain to both of its ends. Figures 2i-l show that this possibility is provided by chains of $Co^{hcp}$ atoms. The spectral intensity of these chains does not show any change when the tip moves along a line starting from the substrate and then across the entire Co chain. This implies, that a close-packed linear chain made from the initially nonmagnetic individual $Co^{hcp}$ atoms on the Re(0001) surface[23] still has a completely quenched magnetization. The superconductivity from the substrate can, thus, penetrate this chain of atoms. Because Fe and Co atoms both occupy hcp adsorption sites, and because of their identical orbital structure on Re, the Co chain might represent an ideal termination for the Fe chain where the latter shows signatures of a topological superconductor.

To follow the idea of terminating the magnetic Fe chain by the nonmagnetic Co chain, we first investigate the magnetic properties at the material transition in hybrid $Co^{hcp}$-$Fe^{hcp}$ chains in the normal metallic state of the substrate. This is done via *ab-initio* calculations using the Korringa-Kohn-Rostoker (KKR) Green's function method based on an embedding scheme together with an effective spin model (see Methods and Supplementary Note 3). Our calculations reveal that the exchange interactions between nearest and next nearest neighbors within the $Fe_{20}$ chain are strongly antiferromagnetic (Supplementary Fig. 3), which leads to spin-frustration. This frustration is resolved by the formation of a cycloidal spin spiral of wavelength between three and four lattice constants (Fig. 3), in agreement with previous experimental results[11]. Consequently, in this system, spin spirals originate from spin-frustration rather than from Dzyaloshinskii-Moriya (DM) interaction. The DM interaction only sets the plane of rotation, which is at an angle of 30° to the surface plane for the $Fe_{20}$ chain (Fig. 3c), and the rotational sense, but it has only a minor effect on the spin-spiral wavelength. Note, that, for the pure $Fe_{20}$ chain, the magnetic moments and interactions of the Fe atoms at both ends differ from the Fe atoms in the interior of the chain (Fig. 3d and Supplementary Figs. 3,4).

However, when terminating the $Fe_{20}$ chain with $Co_5$ chains, both, the magnetic moments and the interactions of these Fe atoms become similar to those of the Fe atoms in the interior of the chain (Fig. 3b,d and Supplementary Figs. 3,4). Moreover, the magnetic moments of the Co atoms in the $Co_5$ chain attached to the $Fe_{20}$ chain are essentially zero. Only the first Co atom at the transition to the Fe chain has a considerable induced magnetic moment (Figs. 3b,d). As a result, the impact of the Co chains on the magnetic structure in the interior of the Fe chain is negligible. Therefore, already five atom long Co chains realize a perfect termination of the $Fe_{20}$ chain with an atomically sharp transition between the Fe chain's spin spiral state and the nonmagnetic $d$-states of the Co chain.

Keeping these results in mind, we experimentally investigate the in-gap electronic structure in the superconducting state of the substrate along hybrid $Fe_{20}$-$Co_5$ and $Co_5$-$Fe_{20}$-$Co_5$ chains which have been built by successively attaching Co atoms first to the right (Fig. 4c,d) and then to the left side (Fig. 4e,f) of the pure $Fe_{20}$ chain (Fig. 4a,b). Indeed, the in-gap electronic structure measured on the last few Fe atoms close to the Co termination is considerably different from those measured on the last few Fe atoms at the open ends in the hybrid chains and in the pure $Fe_{20}$ chain. In particular, the spectral intensity of the YSR band at +0.1 meV, which is increased at the open ends of the pure $Fe_{20}$ chain and at the open end of the $Fe_{20}$-$Co_5$ chain (see arrows in Figs. 4b,d), is almost completely moved out of the gap region, both at the single and at the two Co-terminated ends of the $Fe_{20}$-$Co_5$ and $Co_5$-$Fe_{20}$-$Co_5$ chains, respectively (see also the spectra in Figs. 4g,h for comparison with Figs. 2g,h). Most notably, the zero-energy spectral weight maximum is persistent at the Co-terminated Fe chains. Its position is only slightly shifted towards the interior of the Fe part of the hybrid chain by about two atomic lattice constants after having attached the Co termination. This is visible in the zero-energy spectral weight extracted from Fig. 4b,d,f as shown in Fig. 4i. Simultaneously, the five local maxima and minima of the oscillations of the zero-energy spectral weight in the interior of the chain shift slightly towards the center. This experimental observation is consistent with the interpretation of the zero-energy spectral weight as a signature of a Majorana bound state localized at each end of the $Fe_{20}$ chain, which is expected to be protected against the local perturbation of the $Fe_{20}$ chain by the Co termination. Such a perturbation, which is not affecting the internal spin structure of the $Fe_{20}$ chain, can merely shift the lateral position of a Majorana bound state, but cannot completely remove it. In contrast, it can strongly influence the topologically trivial YSR bands. In order to further support this interpretation, we also investigated the topological properties of infinite Fe chains and the spatially resolved in-gap electronic structure of the pure and Co terminated chains with a tight-binding model using the parameters extracted from the above KKR calculations (see Methods and Supplementary Notes 4,5). For an appropriately tuned superconducting energy-gap $\Delta$, the tight-binding model reproduces the zero-energy spectral weight localized at both ends of the pure Fe chains, whose spatial localization is slightly increased when attaching the Co terminations, as shown in Fig. 4j. Using the same $\Delta$, the tight-binding model for the infinite Fe chain displays the topologically superconducting phase. These results corroborate that the experimentally observed zero-energy spectral weight localized at the ends of the Fe chain is a signature of a Majorana bound state which persists when terminating the chain with topologically trivial Co chains.

Our results, thus, suggest that appropriate terminations of topologically superconducting chains realized by artificial hybrid transition-metal atom chains can be used to tune the properties of Majorana bound states and trivial YSR bands. We thereby establish essential next steps towards the atom-by-atom design of hybrid networks of spin-chains and nonmagnetic superconducting chains and towards the controlled manipulation of Majorana bound states, which are desired for Majorana braiding and the demonstration of topological quantum computation.

## Methods

**Experimental procedures**
All measurements were performed in a home-built ultra-high-vacuum scanning tunneling microscope (STM) setup at $T = 0.3$ K [24]. We used electrochemically etched tungsten tips that were flashed to $T = 1500$ K before inserting them into the STM. The Re(0001) crystal was cleaned by Ar ion sputtering, followed by multiple cycles of $O_2$ annealing at $T = 1530$ K and flashing to $T = 1800$ K. Mn, Fe and Co atoms were successively deposited keeping the substrate at $T < 10$ K. The bias-dependent differential tunneling conductance d$I$/d$V$ was measured using a Lock-In amplifier by modulating the bias voltage $V$ with $V_{mod}$ = 20-40 µV at a frequency of $f_{mod}$ = 4142 kHz, and at constant tip height stabilized at a bias voltage $V_{stab}$ and tunnel current $I_{stab}$ before opening the feedback loop for measurement. The bias voltage is applied to the sample and zero bias corresponds to $E_F$. Single atoms were manipulated using STM-tip-induced atom manipulation by lowering the bias voltage and increasing the setpoint current to the manipulation parameters $V$ = 1 mV and $I$ = 100 nA.

**Ab-initio calculations**
The density functional theory (DFT) calculations were performed employing the full-potential Korringa-Kohn-Rostoker (KKR) Green function method with spin-orbit-coupling added to the scalar relativistic approximation (see Supplementary Note 3)[25]. The exchange and correlation potential is treated within in the local spin density approximation using the parametrization of Vosko, Wilk and Nusair[26]. The Re(0001) substrate is modeled by 22 layers of Re augmented by two vacuum regions corresponding to four interlayer distances. A k-mesh of 150 x 150 and an angular momentum cutoff for the scattering problem of $l_{max}$ = 3 are used. The magnetic chains are deposited in the hcp-stacking position on the Re(0001) surface with a relaxation of 20% of the inter-layer distance towards the Re surface, using an embedding technique. Three systems are investigated: In addition to an $Fe_{20}$ chain we added 5 Co atoms to one end of the chain ($Fe_{20}$-$Co_5$) and 5 Co atoms to each end of the chain ($Co_5$-$Fe_{20}$-$Co_5$). The real-space clusters which are embedded on the Re(0001) surface contain the nearest-neighbor Re atoms (and vacuum sites), resulting in cluster sizes of 146, 181, and 216 sites, respectively. The magnetic exchange interactions were obtained using the magnetic force theorem in the frozen potential approximation and the infinitesimal rotation method[27,28]. The site-resolved magnetic on-site anisotropy is obtained using the method of constraining fields[29].

**Effective spin model calculations**
The magnetic exchange interactions and the on-site magnetic anisotropy obtained from the ab-initio calculations are used to parametrize the following classical Heisenberg model (see Supplementary Note 3),

$$\mathcal{H} = \sum_i \boldsymbol{e}_i \bar{\bar{\mathcal{K}}}_i \boldsymbol{e}_i + \frac{1}{2}\sum_{ij} J_{ij} \boldsymbol{e}_i \cdot \boldsymbol{e}_j + \frac{1}{2}\sum_{ij} \boldsymbol{D}_{ij} \cdot (\boldsymbol{e}_i \times \boldsymbol{e}_j) + \frac{1}{2}\sum_{ij} \boldsymbol{e}_i \bar{\bar{J}}_{ij}^{sym} \boldsymbol{e}_j ,$$

where the unit vectors $\boldsymbol{e}_i$ point along the magnetization of atom $i$ in the chain, $\bar{\bar{\mathcal{K}}}_i$ are the magnetic on-site anisotropy matrices, $J_{ij}$ are the isotropic exchange interactions, $\boldsymbol{D}_{ij}$ are the Dzyaloshinskii-Moriya interaction vectors, and $\bar{\bar{J}}_{ij}^{sym}$ are the symmetric anisotropic exchange interaction matrices. The magnetic ground states are obtained from numerically minimizing the Heisenberg model starting from several random initial configurations of $\boldsymbol{e}_i$.

**Tight-binding model calculations**
The finite magnetic chains on superconducting Re(0001) are modeled by a tight-binding Hamiltonian with realistic parameters obtained from DFT (see Supplementary Notes 4,5). The model explicitly considers the $d$ orbitals of the chain atoms, with parameters that account for the Re substrate via an effective Hamiltonian construction. The site-resolved on-site electronic structure is modeled by a chemical potential, a spin splitting generating the magnetic moments, a local spin-orbit coupling, and an orbital-dependent crystal-field splitting. The hoppings between the sites are assumed to be hermitian, spin-independent and symmetric in the orbitals. Superconductivity is added in the *s*-wave approximation with the local pairing potential being an orbital-independent parameter[30]. The local density of states is obtained by assuming an artificial temperature broadening, which is half of the spectral gap of the superconducting state.

**Data availability**
The authors declare that the data supporting the findings of this study are available within the paper and its supplementary information files.

## Acknowledgements
L.S., M.S., T.P., R.W., and J.W. gratefully acknowledge funding by the Cluster of Excellence 'Advanced Imaging of Matter' (EXC 2056 - project ID 390715994) of the Deutsche Forschungsgemeinschaft (DFG). L.S., M.S., R.W., and J.W. acknowledge support by the SFB 925 'Light induced dynamics and control of correlated quantum systems' of the Deutsche Forschungsgemeinschaft (DFG). L.S. and R.W. acknowledge funding by the ERC Advanced Grant ADMIRE (No. 786020). S.B., M.d.S.D. and S.L. acknowledge funding from the European Research Council (ERC) under the European Union's Horizon 2020 research and innovation program (ERC-consolidator grant 681405—DYNASORE) and the computing time granted by the JARA-HPC Vergabegremium and VSR commission on the supercomputer JURECA at Forschungszentrum Jülich.

## Competing interests
The authors declare no competing interests.

## Author contributions
L.S. and J.W. conceived the experiments. L.S. did the measurements. M.S. and J.H. started the initial experiments together with L.S.. L.S. and J.W. analyzed the experimental data. S.B. performed the KKR-based DFT calculations, the effective spin model calculations, and the tight-binding model simulations. T.P. contributed to the tight-binding model simulations. S.B., M.d.S.D. and S.L. conceived the theoretical framework and analyzed the results of all calculations. L.S. prepared the figures and J.W. wrote the paper. All authors contributed to the discussions and to correcting the manuscript.

## References

1. Choy, T. P., Edge, J. M., Akhmerov, A. R. & Beenakker, C. W. J. Majorana fermions emerging from magnetic nanoparticles on a superconductor without spin-orbit coupling. *Phys. Rev. B* **84**, 195442 (2011).
2. Martin, I. & Morpurgo, A. F. Majorana fermions in superconducting helical magnets. *Phys. Rev. B* **85**, 144505 (2012).
3. Klinovaja, J., Stano, P., Yazdani, A. & Loss, D. Topological superconductivity and Majorana fermions in RKKY systems. *Phys. Rev. Lett.* **111**, 186805 (2013).
4. Nadj-Perge, S., Drozdov, I. K., Bernevig, B. A. & Yazdani, A. Proposal for realizing Majorana fermions in chains of magnetic atoms on a superconductor. *Phys. Rev. B* **88**, 020407(R) (2013).
5. Pientka, F., Glazman, L. I. & von Oppen, F. Topological superconducting phase in helical Shiba chains. *Phys. Rev. B* **88**, 155420 (2013).
6. Vazifeh, M. M. & Franz, M. Self-Organized Topological State with Majorana Fermions. *Phys. Rev. Lett.* **111**, 206802 (2013).
7. Braunecker, B. & Simon, P. Interplay between classical magnetic moments and superconductivity in quantum one-dimensional conductors: Toward a self-sustained topological Majorana phase. *Phys. Rev. Lett.* **111**, 147202 (2013).
8. Nadj-Perge, S. *et al.* Observation of Majorana fermions in ferromagnetic atomic chains on a superconductor. *Science* **346**, 602 (2014).
9. Ruby, M. *et al.* End States and Subgap Structure in Proximity-Coupled Chains of Magnetic Adatoms. *Phys. Rev. Lett.* **115**, 197204 (2015).

10. Pawlak, R. *et al.* Probing Atomic Structure and Majorana Wavefunctions in Mono-Atomic Fe-chains on Superconducting Pb-Surface. *npj Quantum Information* **2**, 16035 (2016).
11. Kim, H. *et al.* Toward tailoring Majorana bound states in artificially constructed magnetic atom chains on elemental superconductors. *Sci. Adv.* **4**: eaar5251 11 (2018).
12. Feldman, B. E. *et al.* High-resolution studies of the Majorana atomic chain platform. *Nat. Phys.* **13**, 286 (2017).
13. Jeon, S. *et al.* Distinguishing a Majorana zero mode using spin-resolved measurements. *Science* 10.1126/science.aan3670 (2017).
14. Ruby, M., Heinrich, B. W., Peng, Y., von Oppen, F. & Franke, K. J. Exploring a Proximity-Coupled Co Chain on Pb(110) as a Possible Majorana Platform. *Nano Lett.* **17**, 4473 (2017).
15. Kamlapure, A., Cornils, L., Wiebe, J. & Wiesendanger, R. Engineering the spin couplings in atomically crafted spin chains on an elemental superconductor. *Nat. Commun.* **9**, 3253 (2018).
16. Theiler, A., Björnson, K. & Black-Schaffer, A. M. Majorana bound state localization and energy oscillations for magnetic impurity chains on conventional superconductors. *Phys. Rev. B* **100**, 214504 (2019).
17. Choi, D.-J. *et al.* Colloquium: Atomic spin chains on surfaces. *Rev. Mod. Phys.* **91**, 041001 (2019).
18. Kitaev, A. Unpaired Majorana fermions in quantum wires. *Phys. Usp.* **44**, 131 (2001).
19. Alicea, J. New directions in the pursuit of Majorana fermions in solid state systems. *Reports Prog. Phys.* **75**, 76501 (2012).
20. Lazarovits, B., Szunyogh, L. & Weinberger, P. Magnetic properties of finite Co chains on Pt(111). *Phys. Rev. B* **67**, 24415 (2003).
21. Cardias, R. *et al.* Magnetic and electronic structure of Mn nanostructures on Ag(111) and Au(111). *Phys. Rev. B* **93**, 014438 (2016).
22. Lászlóffy, A., Rózsa, L., Palotás, K., Udvardi, L. & Szunyogh, L. Magnetic structure of monatomic Fe chains on Re(0001): Emergence of chiral multispin interactions. *Phys. Rev. B* **99**, 184430 (2019).
23. Schneider, L. *et al.* Magnetism and in-gap states of 3d transition metal atoms on superconducting Re. *npj Quantum Materials* **4**, 42 (2019).
24. Wiebe, J. *et al.* A 300 mK ultra-high vacuum scanning tunneling microscope for spin-resolved spectroscopy at high energy resolution. *Rev. Sci. Instrum.* **75**, 4871 (2004).
25. Bauer, D. S. G. Development of a relativistic full-potential first-principles multiple scattering Green function method applied to complex magnetic textures of nano structures at surfaces, Dissertation RWTH Aachen University (2013).
26. Vosko, S. H., Wilk, L., & Nusair, M. Accurate spin-dependent electron liquid correlation energies for local spin density calculations: a critical analysis. *Can. J. Phys.* **58**, 1200 (1980).
27. Liechtenstein, A. I., Katsnelson, M. I., Antropov, V. P. & Gubanov, V. A. Local spin density functional approach to the theory of exchange interactions in ferromagnetic metals and alloys. *J. Magn. Magn. Mater.* **67**, 65 (1987).
28. Ebert, H. & Mankovsky, S. Anisotropic exchange coupling in diluted magnetic semiconductors : Ab initio spin-density functional theory. *Phys. Rev. B* **79**, 045209 (2009).
29. Brinker, S., dos Santos Dias, M. & Lounis, S. The chiral biquadratic pair interaction. *New J. Phys.* **21**, 083015 (2019).
30. Li, J. *et al.* Topological superconductivity induced by ferromagnetic metal chains. *Phys. Rev. B* **90**, 235433 (2014).

# Figures

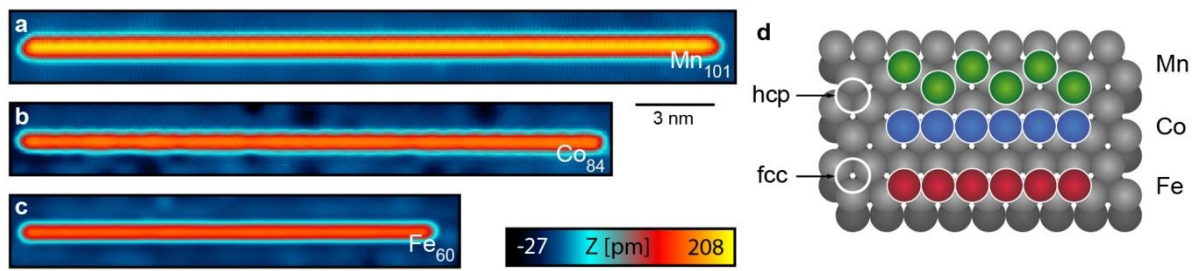

**Figure 1 | Geometric structures of artificial 3d transition metal chains built on a Re(0001) surface.**
**a**, Constant-current STM image of a zig-zag chain consisting of 101 Mn atoms, with adsorption sites alternating between the fcc and hcp hollow sites on the Re substrate. **b**, Constant-current STM image of a linear chain of 84 hcp Co atoms. **c**, Constant-current STM image of a linear chain of 60 hcp Fe atoms. All data (**a-c**) is recorded with $V$ = 6 mV, $I$ = 0.2 nA. **d**, Sketches of the atomic positions in (**a-c**).

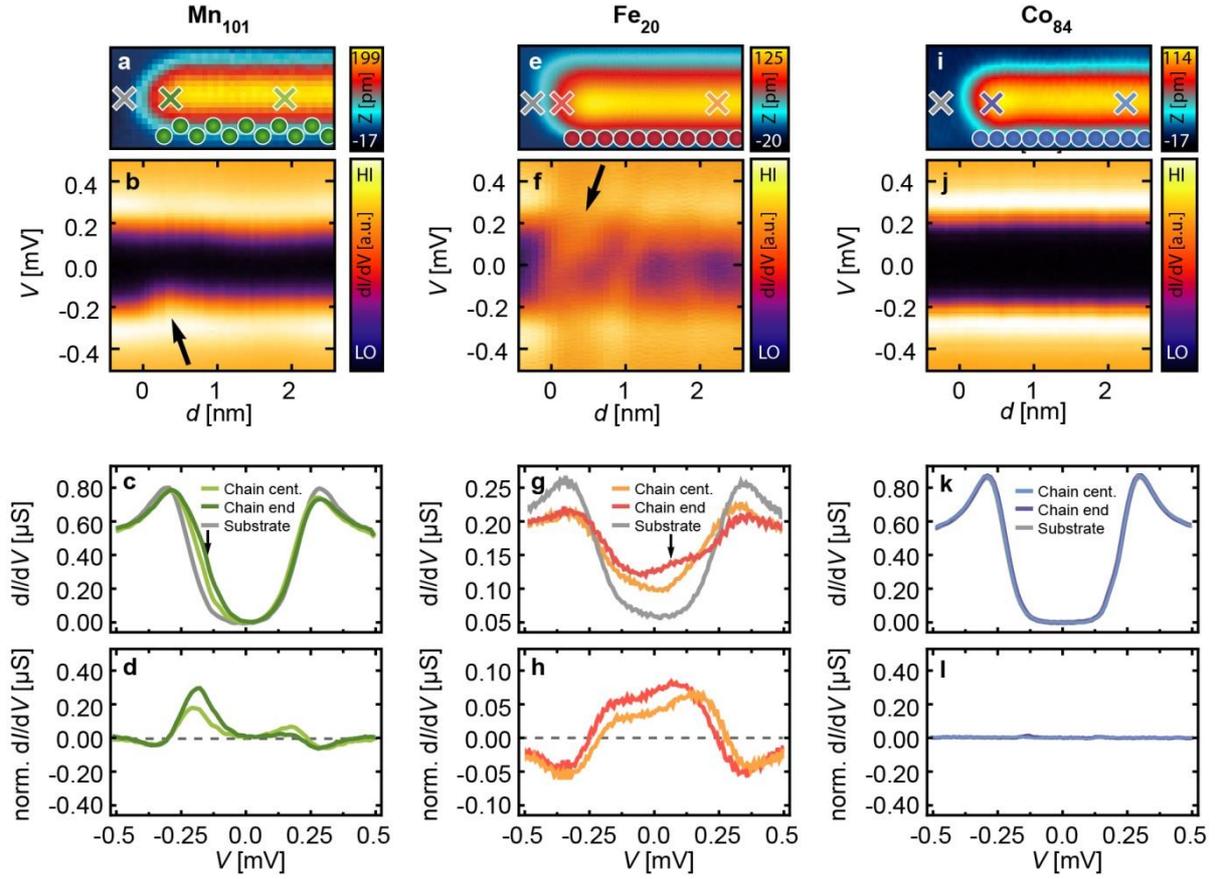

**Figure 2 | Low energy electronic structure at the terminations of the Mn, Fe, and Co chains. a**, Constant-current STM image of the end of a $Mn_{101}$ chain. **b** Differential tunneling conductance along the $Mn_{101}$ chain aligned with the topography shown in (**a**). **c**, d$I$/d$V$ spectra of the $Mn_{101}$ chain and substrate taken at the positions marked by crosses in (**a**) ($V_{stab}$ = 1 mV, $I_{stab}$ = 0.5 nA, $V_{mod}$ = 40 µV). **d**, d$I$/d$V$ spectra from (**c**) normalized by subtraction of the substrate spectrum. **e**, Constant-current STM image of the end of a $Fe_{20}$ chain. **f**, Differential tunneling conductance along the $Fe_{20}$ chain. **g**, d$I$/d$V$ spectra of the $Fe_{20}$ chain and substrate taken at the positions marked by crosses in (**e**) ($V_{stab}$ = 1 mV, $I_{stab}$ = 0.2 nA, $V_{mod}$ = 20 µV). **h**, d$I$/d$V$ spectra from (**g**) normalized by subtraction of the substrate spectrum. **i**, Constant-current STM image of the end of a $Co_{84}$ chain. **j**, Differential tunneling conductance along the $Co_{84}$ chain. **k**, d$I$/d$V$ spectra of the $Co_{84}$ chain and substrate taken at the positions marked by crosses in (**i**) ($V_{stab}$ = 1 mV, $I_{stab}$ = 0.5 nA, $V_{mod}$ = 40 µV). **l**, d$I$/d$V$ spectra from (**k**) normalized by subtraction of the substrate spectrum. All data (**a**, **e**, **i**) is recorded with $V$ = 6 mV, $I$ = 0.2 nA. Arrows in (**b**, **c**, **f**, **g**): see text.

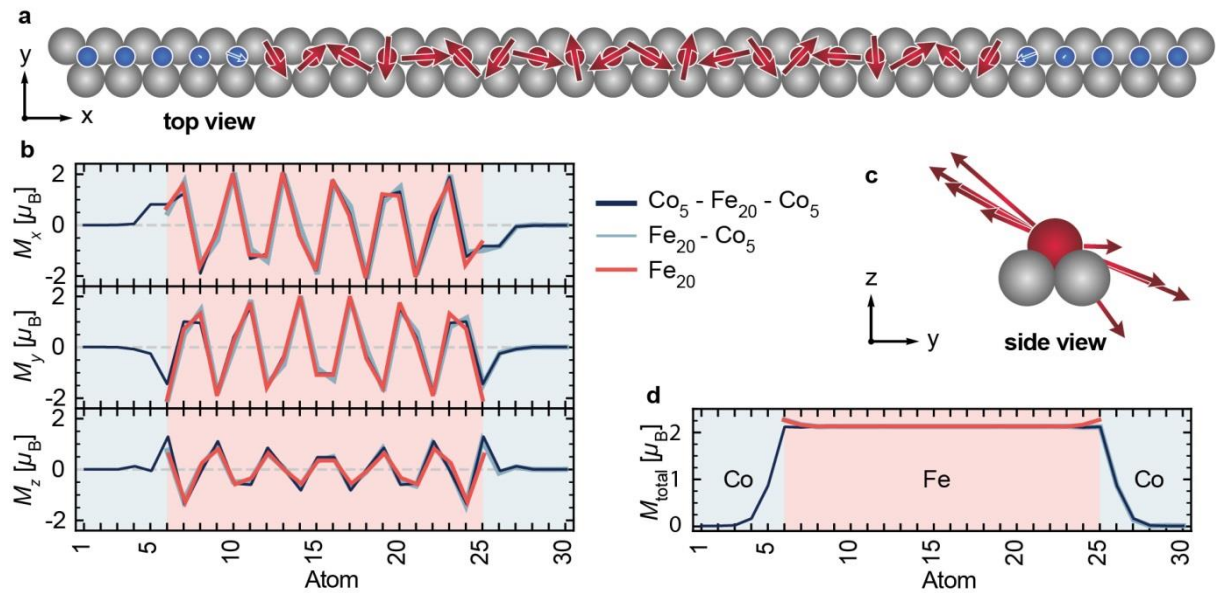

**Figure 3 | Calculated spin structure of the $Fe_{20}$ chain without and with $Co_5$-termination. a**, Top view of the calculated spin structure in the $Co_5$-$Fe_{20}$-$Co_5$ chain. Blue and red spheres correspond to Co and Fe atoms, respectively. The length of the arrows is proportional to the size of the in-plane component of the magnetic moment at each particular site. **b**, x, y and z components of the magnetic moments in the $Co_5$-$Fe_{20}$-$Co_5$-, the $Fe_{20}$-$Co_5$- and the $Fe_{20}$-chains. **c**, Side view of the $Co_5$-$Fe_{20}$-$Co_5$ chain cycloidal spin spiral along the negative $x$ direction. **d**, Total magnetic moment of the atoms along all three chains.

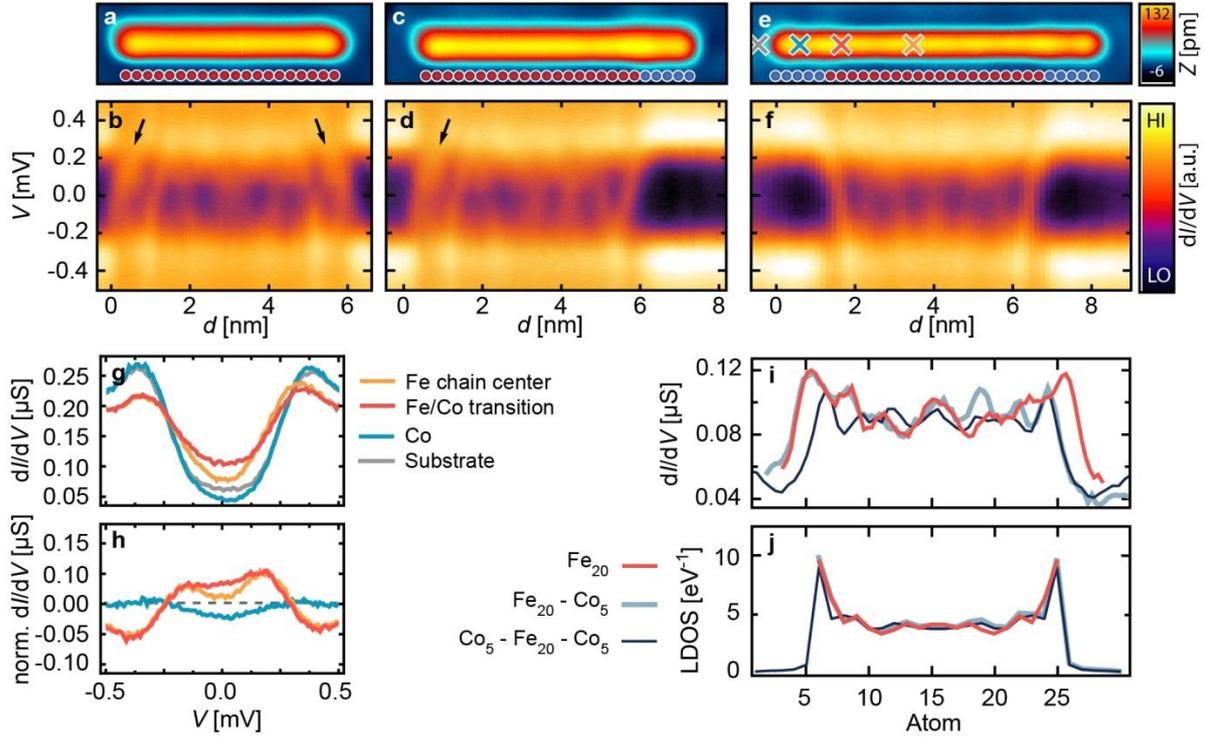

**Figure 4 | In-gap states of the Fe$_{20}$ chains without and with Co$_5$-termination. a**, Constant-current STM image of the Fe$_{20}$ chain. **b**, Differential tunneling conductance along the Fe$_{20}$ chain aligned with the topography in (**a**). **c**, Constant-current STM image of the Fe$_{20}$-Co$_5$ chain. **d**, Differential tunneling conductance along the Fe$_{20}$-Co$_5$ chain. **e**, Constant-current STM image of the Co$_5$-Fe$_{20}$-Co$_5$ chain. **f**, Differential tunneling conductance along the Co$_5$-Fe$_{20}$-Co$_5$ chain. Arrows in (**b**, **d**): see text. **g**, d$I$/d$V$ spectra of the Co$_5$-Fe$_{20}$-Co$_5$ chain taken at the positions marked by crosses in (**e**) ($V_{stab}$ = 1 mV, $I_{stab}$ = 0.2 nA, $V_{mod}$ = 20 µV ). **h**, d$I$/d$V$ spectra from (**g**) normalized by subtraction of the substrate spectrum. **i**, Experimental zero-bias d$I$/d$V$ signal along the three different chains as indicated left of (**j**) extracted from (**b**, **d**, **f**). **j**, Spectral weight at the Fermi energy along the three different chains as indicated on the left, calculated from the tight-binding model. The tight-binding model parameters, except for $\Delta$, are extracted from the *ab-initio* calculations (Methods and Supplementary Notes 4, 5). $\Delta$ is chosen as to reproduce the experimentally observed spatial decay of the zero-bias spectral weight from the termination of the spin chain towards the chain center.